\documentclass[procedia]{easychair}

\usepackage{doc}
\usepackage{makeidx}

\usepackage{cite}

\newcommand{\means}[1]{\langle#1\rangle}

\def\procediaConference{99th Conference on Topics of
  Superb Significance (COOL 2014)}

\title{Phase Diagram of the Kitaev-type Model on a Decorated Honeycomb Lattice in the Isolated Dimer Limit}

\titlerunning{Phase Diagram of the Kitaev-type Model on \ldots}

\author{
    Joji Nasu\inst{1}
\and
    Yukitoshi Motome\inst{2}
}

\institute{
  Tokyo Institute of Technology,
  Ookayama, Meguro, Tokyo, Japan\\
  \email{nasu@phys.titech.ac.jp}
\and
   University of Tokyo,
   Hongo, Bunkyo, Tokyo, Japan\\
   \email{motome@ap.t.u-tokyo.ac.jp}\\
 }

\authorrunning{Nasu and Motome}

\usepackage{amsmath,amssymb}
\begin{document}

\maketitle

\keywords{Kitaev model, chiral spin liquid, toric code, Monte Carlo simulation}

\begin{abstract}
 An effective model in the isolated dimer limit of the Kitaev-type model on a decorated honeycomb lattice is investigated at finite temperature.
The ground state of this model is exactly shown to be a chiral spin liquid with spontaneous breaking of time reversal symmetry.
 We elaborate the finite-temperature phase diagram by using the mean-field approximation and Monte Carlo simulation.
 We find that the phase transition between the high-temperature paramagnetic phase and the low-temperature chiral spin liquid phase is always of second order in the Monte Carlo results, although a tricritical point appears in the mean-field phase diagram. 
 The finite-size scaling analysis of the Monte Carlo data indicates that the phase transition belongs to the two-dimensional Ising universality class.
\end{abstract}


%
%

\section{Introduction}

A Mott insulator without magnetic ordering down to the lowest temperature ($T$) has attracted considerable attention in strongly correlated electron systems~\cite{ISI:000275366100033}.
This state, called the quantum spin liquid, is brought about by strong quantum fluctuations, which often become conspicuous in low-dimensional systems and in the presence of geometrical frustration of the lattice structures.
Several candidates of quantum spin liquids have been found theoretically and experimentally in quantum spin systems on geometrically frustrated lattices.
Among them, the chiral spin liquid (CSL), where the time reversal symmetry is broken despite the absence of magnetic order, is of particularly interesting and has been discussed, for instance, in the Heisenberg models on triangular and kagome lattices~\cite{PhysRevB.39.11879,PhysRevLett.70.2641,PhysRevLett.112.137202,ISI:000341933300006,ISI:000343980700001}.

As a simple realization of CSL, a Kiteav-type model defined on a decorated honeycomb lattice has been studied in the last decade~\cite{Kitaev2006,PhysRevLett.99.247203,PhysRevB.78.125102,PhysRevB.82.174412,PhysRevLett.105.067207,PhysRevB.81.060403,Nasu2015}.
 This is a variant of the Kitaev model on the honeycomb lattice~\cite{Kitaev2006}, for which the ground state is exactly shown to be a CSL.
 The decorated honeycomb lattice is constructed from the honeycomb lattice by replacing the sites by triangles [see Fig.~\ref{fig:lattice}(a)].
 This model exhibits two types of CSLs, topologically trivial (with Abelian excitations) and nontrivial (with non-Abelian excitations), by changing the exchange constants~\cite{PhysRevLett.99.247203}.
 The characteristics of the CSL ground states were also studied for the effective models in two limits, the isolated dimer limit and the isolated triangle limit~\cite{PhysRevB.78.125102}.
Recently, the authors studied the finite-$T$ properties of CSLs for the original Kitaev-type model by  using the quantum Monte Carlo (MC) simulation~\cite{Nasu2015}.
In addition to the finite-$T$ phase transition associated with the time reversal symmetry breaking, peculiar crossovers were found in the regions near the two limiting cases. 
Although the results were compared with the asymptotic behaviors expected in the two limits, the details were not reported in the previous study~\cite{Nasu2015}.
 In order to further understand the mechanism of the crossovers and the phase transition, it is desired to elucidate the finite-$T$ properties of the effective models in these limiting cases.

In this paper, we investigate thermodynamic properties in the isolated dimer limit of the Kitaev-type model on the decorated honeycomb lattice.
In this limit, the effective model is described by Ising-type $Z_2$ variables only.
Performing the mean-field (MF) approximation and an unbiased MC simulation, we elaborate the finite-$T$ phase diagram.
In both calculations, we find that the system exhibits a phase transition associated with the time reversal symmetry breaking at finite $T$.
The phase transition is always of second order in the MC simulation, whereas both first- and second-order transitions appear with a tricritical point in the MF approximation.
We also find a crossover above the critical temperature, associated with the coherent alignment of local conserved quantities.
Moreover, the analysis of the finite-size scaling indicates that the phase transition belongs to the two-dimensional Ising universality class.


\section{Model}\label{sec:model}

\begin{figure}[tb]
 \begin{centering}
  \includegraphics[width=0.9\textwidth]{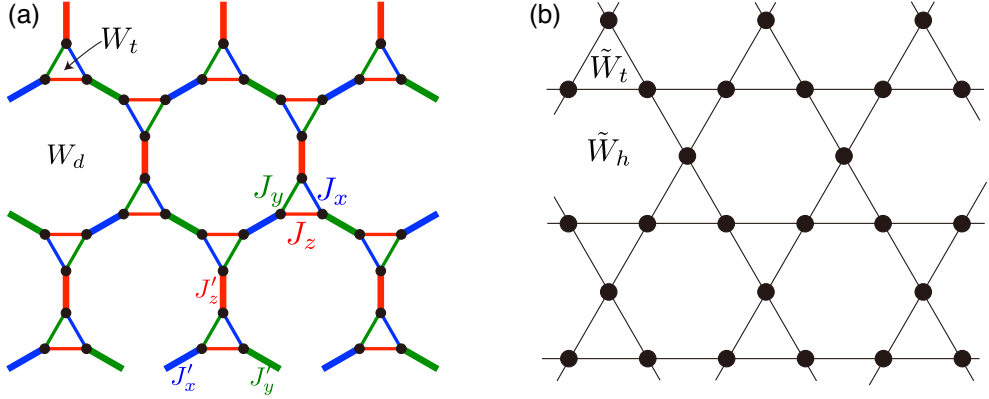}
  \caption{(a) Schematic picture of the model in Eq.~(\ref{eq:1}) on a decorated honeycomb lattice.
  Thin lines represent NN $\gamma$ bonds in triangles and thick lines represent NN $\gamma'$ bonds connecting triangles.
(b) A kagome lattice on which the effective model in Eq.~(\ref{eq:2}) is defined.
  }
  \label{fig:lattice}
 \end{centering}
\end{figure}

We begin with a variant of the Kitaev model defined on a decorated honeycomb lattice depicted in Fig.~\ref{fig:lattice}(a), whose Hamiltonian is given by~\cite{PhysRevLett.99.247203}
\begin{align}
 {\cal H}=-\sum_{\gamma =x,y,z}\sum_{\means{ij}_\gamma}J_\gamma \sigma_i^\gamma \sigma_j^\gamma
-\sum_{\gamma =x,y,z}\sum_{\means{ij}'_\gamma}J'_\gamma \sigma_i^\gamma \sigma_j^\gamma,\label{eq:1}
\end{align}
where $\sigma_i^\gamma$ is the $\gamma(=x,y,z)$ component of the Pauli matrices, representing an $S=1/2$ spin located on site $i$;
$\means{ij}_\gamma$ and $\means{ij}_\gamma'$ stand for the nearest neighbor (NN) pairs on $\gamma$ bonds within the triangles and on $\gamma'$ bonds connecting the triangles, respectively [see Fig.~\ref{fig:lattice}(a)].
For simplicity, we take $J_x=J_y=J_z=J$ and $J_x'=J_y'=J_z'=J'$, as in Ref.~\cite{PhysRevLett.99.247203}.

There are two kinds of conserved quantities in this model: $W_d=\prod_{i\in d}\sigma_i^{\gamma_i}$ and $W_t=\prod_{i\in t}\sigma_i^{\gamma_i}$, where $d$ ($t$) represents a set of the sites belonging to a dodecagon (triangle) loop on the decorated honeycomb lattice, and $\gamma_i$ corresponds to the bond component not included in the loop $d$ or $t$ among three NN bonds connected to the site $i$.
These conserved quantities are $Z_2$ variables taking the values $\pm1$.
Note that $W_t$ changes its sign by the time reversal operation, as it consists of the product of three spins.
The ground state of this model is a CSL in which the time reversal symmetry is broken by uniform alignment of $W_t$ as well as $W_d$~\cite{PhysRevLett.99.247203}.

In the present study, we are particularly interested in the limit of $J'\gg J$, where the dimers on the inter-triangle $\gamma'$ bonds are weakly coupled by $J$. 
This is called the isolated dimer limit.
When each dimer is regarded as a site, the interacting dimers form a kagome lattice, as shown in Fig.~\ref{fig:lattice}(b).
The effective interaction between the dimers is derived by the perturbation expansion in terms of $J/J'$~\cite{PhysRevB.78.125102}.
The effective Hamiltonian is written in the form of
\begin{align}
 {\cal H}_{\rm eff}=E_0 -C_h \sum_h \tilde{W}_h-C_t\sum_{\means{h t_1 t_2}}\tilde{W}_h\tilde{W}_{t_1}\tilde{W}_{t_2},\label{eq:2}
\end{align}
where $\tilde{W}_h$ and $\tilde{W}_t$ are conserved quantities given by the projection of $W_d$ and $W_t$ onto the dimer subspace; they are also $Z_2$ variables taking $\pm 1$.
As shown in Fig.~\ref{fig:lattice}(b), $\tilde{W}_h$ and $\tilde{W}_t$ are defined on hexagons and triangles, respectively.
Similar to $W_t$, $\tilde{W}_t$ also changes its sign by the time reversal operation.
The coefficients in Eq.~(\ref{eq:2}) are given by $C_h=\frac{63}{256}\frac{J^6}{J'^5}-\frac{297}{1024}\frac{J^8}{J'^7}$ and $C_t=\frac{33}{2048}\frac{J^8}{J'^7}$;
$E_0$ is a constant, and the sum of $\means{h t_1 t_2}$ is taken over three neighboring plaquettes~\cite{PhysRevB.78.125102}.

In the following calculations, we parameterize the coefficients in Eq.~(\ref{eq:2}) as $C_h=x$ and $C_t=(1-x)/15$ by introducing $x \in [0:1]$.
The limit of $J'\gg J$ corresponds to $x\to 1$, where the second term in Eq.~(\ref{eq:2}) is dominant over the third term.
In the present study, however, we analyze the effective model in Eq.~(\ref{eq:2}) in the whole region of $x$, i.e., $0\le x\le 1$, in order to elucidate the thermodynamic properties in this model.

\section{Method}\label{sec:method}

We investigate thermodynamic properties of the effective model in Eq.~(\ref{eq:2}) using the MF approximation and the MC simulation.
In the MF approximation, the interaction between $\tilde{W}_h$ and $\tilde{W}_t$ in the last term in Eq.~(\ref{eq:2}) is decoupled by introducing the mean fields $\means{\tilde{W}_h}$ and $\means{\tilde{W}_t}$. 
The resultant MF Hamiltonian per unit cell is given by
\begin{align}
 {\cal H}_{\rm eff}^{\rm MF}=-(x+(1-x)\means{\tilde{W}_t}^2)W_h-2(1-x)\means{\tilde{W}_h}\means{\tilde{W}_t}\tilde{W}_t+2(1-x)\means{\tilde{W}_h}\means{\tilde{W}_t}^2,
\end{align}
up to a constant.
From this form, the self-consistent equations for the mean fields are given by
\begin{align}
 \means{\tilde{W}_h}&=\tanh[ \beta\{x+(1-x)\means{\tilde{W}_t}^2\}],\\
\means{\tilde{W}_t}&=\tanh\{2\beta (1-x)\means{\tilde{W}_h}\means{\tilde{W}_t}\},
\end{align}
where $\beta=1/T$ is the inverse temperature (we set the Boltzmann constant $k_{\rm B}=1$). 
By solving the above equations and minimizing the free energy, we calculated the thermal averages $\means{\tilde{W}_h}$ and $\means{\tilde{W}_t}$.

In the MC simulation, one can apply a classical MC method as the model in Eq.~(\ref{eq:2}) includes only the Ising type variables $\tilde{W}_h$ and $\tilde{W}_t$. 
We adopted the Wang-Landau algorithm~\cite{PhysRevLett.86.2050,PhysRevE.64.056101}: after obtaining the density of states iteratively, we spent $10^6$ MC steps for measurement and performed them 10 times, independently.
We carried out the simulations for the finite-size clusters including $N=3L^2$ of $\tilde{W}_h$ and $\tilde{W}_t$ up to $L=50$ under the periodic boundary conditions.

\section{Results}\label{sec:results}

\subsection{Mean-field calculation}

\begin{figure}[tb]
 \begin{centering}
  \includegraphics[width=0.8\textwidth]{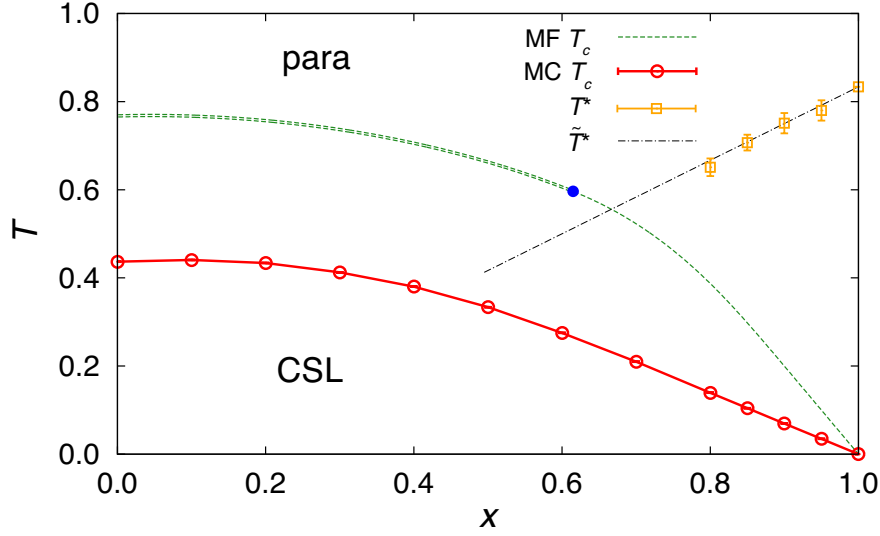}
  \caption{Phase diagram of the effective model in Eq.~(\ref{eq:2}).
  The horizontal axis $x$ is defined so that $C_h = x$ and $C_t = (1-x)/15$ in Eq.~(\ref{eq:2}).
  The single- and double-dashed green lines represent the critical temperature $T_c$ for the second- and first-order phase transition obtained by the MF calculation; the blue filled circle represents the tricritical point.
  The open circles connected by the solid red line and the open squares represent $T_c$ and $T^*$ obtained by the MC simulation,  respectively.
  The dashed-dotted line indicates a crossover temperature $\tilde{T}^*$ estimated for an effective two-level system.
  See the text for details.
  }
  \label{fig:phase}
 \end{centering}
\end{figure}

The phase diagram obtained by the MF calculations is shown in Fig.~\ref{fig:phase}.
There is a phase transition between two phases, the high-$T$ paramagnetic phase and the low-$T$ CSL phase, associated with the time reversal symmetry breaking.
The phase boundary is calculated as follows.
As the CSL phase is characterized by the chiral order parameter $\means{\tilde{W}_t}$, let us consider the expansion of the free energy per unit cell in terms of $\means{\tilde{W}_t}$ as
\begin{align}
 F^{\rm MF}=-\frac{1}{\beta}\ln {\rm Tr} e^{-\beta {\cal H}_{\rm eff}^{\rm MF}}=f_0+f_2\means{\tilde{W}_t}^2 +f_4\means{\tilde{W}_t}^4 +\cdots,
\end{align}
where the coefficients are given by
\begin{align}
 f_2&=2(1-x) \{1-2\beta (1-x)\tanh\beta x \}\tanh\beta x,\\
f_4&=4\beta (1-x)^2[3\{3-8\beta(1-x)\tanh \beta x\}(1-\tanh^2\beta x)+ 8\beta^2(1-x)^2\tanh^4\beta x].
\end{align}
When the transition is of second order, the critical temperature $T_c$ is determined by $f_2=0$ with $f_4>0$, which leads to the equation $1=2\beta(1-x)\tanh\beta x$. 
We find that this condition is indeed met in the large $x$ region; the obtained $T_c$ is shown by the single-dashed green line in Fig.~\ref{fig:phase}.
In the vicinity of $x=1$, $T_c$ arises linearly from zero, as the interaction coefficient $C_t$ in Eq.~(\ref{eq:2}) is proportional to $(1-x)$.
While decreasing $x$, $f_4$ decreases to zero, which results in the tricritical point with $f_2=f_4=0$ at $(x_{\rm tri}, T_{\rm tri})\simeq (0.615, 0.596)$, as shown in Fig.~\ref{fig:phase}.
For $x < x_{\rm tri}$, the phase transition becomes first order; in this region, $T_c$ is determined by comparing two local minima of $F^{\rm MF}$ at $\means{\tilde{W}_t}=0$ and $\means{\tilde{W}_t}>0$.
The obtained value of $T_c$ for the first-order phase transition is presented by the double-dashed green line in Fig.~\ref{fig:phase}.

\subsection{Monte Carlo simulation}

\begin{figure}[tb]
 \begin{centering}
  \includegraphics[width=0.9\textwidth]{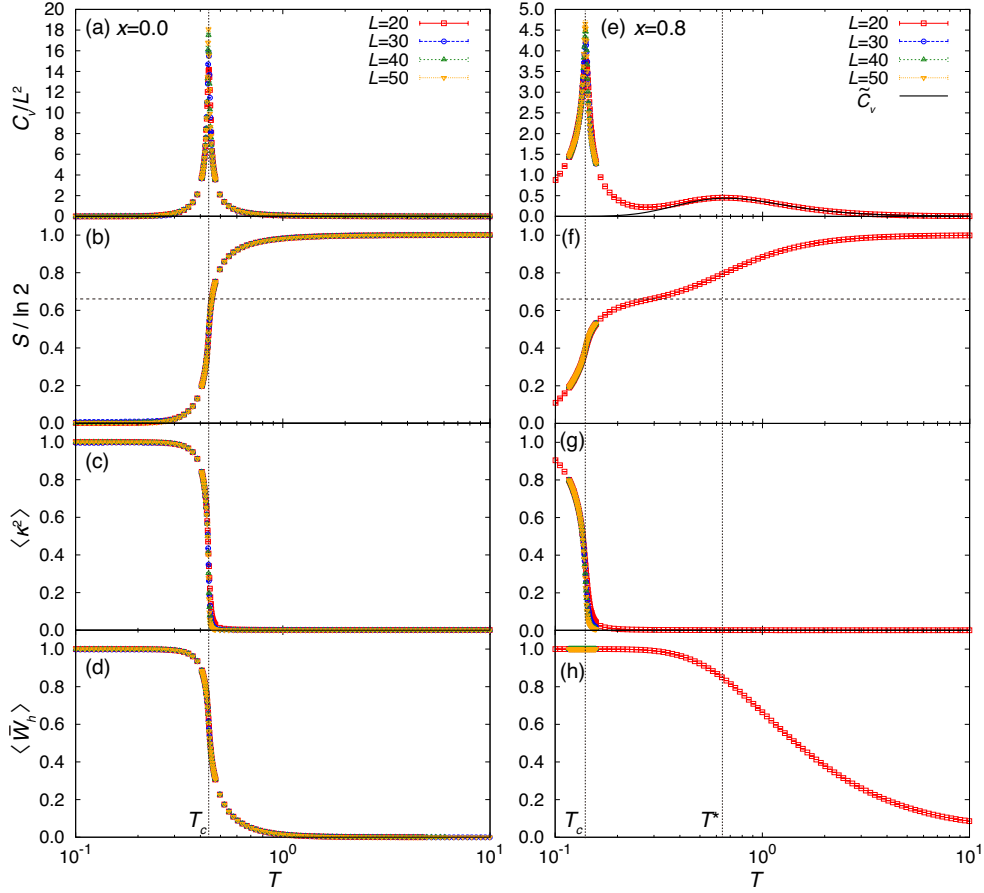}
  \caption{(a) $T$ dependences of (a) the specific heat, (b) the entropy per site, (c) the mean square of the chiral order parameter $\kappa$, and (d) the mean of $\tilde{W}_h$ per hexagon at $x=0.0$ obtained by the MC simulation.
  (e)-(f) Corresponding data at $x=0.8$.
 For comparison, the specific heat for an effective two-level model, $\tilde{C}_v$, is also shown in (e).
  The vertical dotted lines indicate $T_c$ and $T^*$.
  The horizontal dotted lines in (b) and (f) represent $S=\frac23 \ln 2$.
  }
  \label{fig:phys}
 \end{centering}
\end{figure}

We further examined the finite-$T$ properties by the MC simulation, in which the effect of thermal fluctuations is taken into account beyond the MF approximation. 
Figure~\ref{fig:phys} show typical MC results for $T$ dependences of physical quantities.
First, we discuss the data at $x=0$.
Figure~\ref{fig:phys}(a) shows the specific heat for several cluster sizes.
There is a peak that becomes sharper for larger $L$.
The result indicates that the system exhibits a phase transition at $T_c \simeq 0.44$.
Through this transition, all of the entropy $\ln 2$ is released, as shown in Fig.~\ref{fig:phys}(b).
Indeed, both $\tilde{W}_t$ and $\tilde{W}_h$ grow coherently at $T_c$. 
As shown in Fig.~\ref{fig:phys}(c), the mean square of $\tilde{W}_t$, $\means{\kappa^2}$, where $\kappa=\frac{1}{2L^2}\sum_t\tilde{W}_t$, abruptly increases below $T_c$ with decreasing $T$, indicating that the phase transition is associated with the time reversal symmetry breaking.
At the same time, the mean of $\tilde{W}_h$ per hexagon, $\means{\bar{W}_h}=\frac{1}{L^2}\sum_h\means{\tilde{W}_h}$, also increases rapidly around $T_c$, as shown in Fig.~\ref{fig:phys}(d).

Next, we discuss the data at $x=0.8$.
In contrast to the case of $x=0$, the specific heat shows two peaks, as shown in Fig.~\ref{fig:phys}(e).
The low-$T$ peak is similar to the one in Fig.~\ref{fig:phys}(a), which indicates a phase transition. 
On the other hand, the high-$T$ peak is broad and does not show the system size dependence.
This corresponds to a crossover, and we denote the peak temperature by $T^*$.
Figure~\ref{fig:phys}(f) shows the $T$ dependence of the entropy per site. 
The entropy is released successively as decreasing $T$: 
one third of the entropy, $\frac13 \ln 2$, is released through the crossover at $T^*$, and the rest two third $\frac23 \ln 2$ is released through the phase transition at $T_c$.
The latter is associated with $W_t$, as indicated by the growth of $\means{\kappa^2}$ shown in Fig.~\ref{fig:phys}(g).
On the other hand, the former comes from $\tilde{W}_h$; indeed, as shown in Fig.~\ref{fig:phys}(h), $\means{\bar{W}_h}$ increases around $T^*$ with decreasing $T$.
Therefore, $T^*$ corresponds to the coherent growth of $\tilde{W}_h$, while $T_c$ corresponds to the time reversal symmetry breaking signaled by the abrupt growth of $\tilde{W}_t$.
Thus, two conserved quantities develop at different $T$ scales at $x=0.8$, in contrast to the results at $x=0$, where both grow simultaneously near $T_c$.

While the phase transition is driven by the last term in Eq.~(\ref{eq:2}), the crossover at $T^*$ is caused by the second term. 
This is confirmed by considering the second term only, which gives an effective two-level system with the energy splitting $2x$.
The specific heat for this system is given by $\tilde{C}_v=\beta^2 x^2/\cosh^2\beta x$.
The result at $x=0.8$ is presented by the black curve in Fig.~\ref{fig:phys}(e).
This agrees with the MC result near and above $T^*$.

\begin{figure}[tb]
 \begin{centering}
  \includegraphics[width=0.9\textwidth]{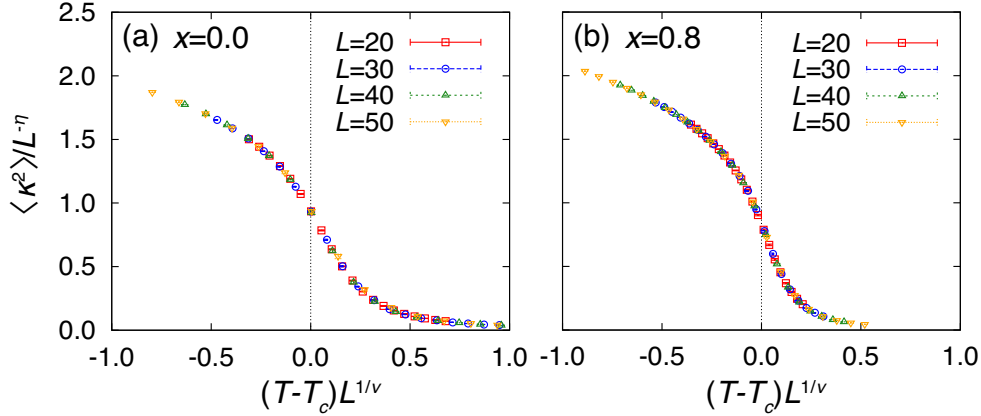}
  \caption{Scaling collapses for $\means{\kappa^2}$ at (a) $x=0.0$ with $T_c=0.4367$, $1/\nu=1.027$, and $\eta=0.243$ at $x=0.0$ and (b) $x=0.8$ with $T_c=0.1390$, $1/\nu=1.001$, and $\eta=0.2545$.}
  \label{fig:fss}
 \end{centering}
\end{figure}

To determine $T_c$ precisely and to clarify the nature of the phase transition, we performed the finite-size scaling for the order parameter $\means{\kappa^2}$.
Figures~\ref{fig:fss}(a) and~\ref{fig:fss}(b) show the scaling collapses of $\means{\kappa^2}$ at $x=0$ and $x=0.8$, respectively.
In both cases, all the data well collapse onto a single universal curve, which indicates that the transitions are of second order.
Using the Bayesian scaling analysis for the optimization~\cite{PhysRevE.84.056704}, we obtain the critical temperature and the critical exponents as
$T_c=0.4367(1)$, $1/\nu=1.027(9)$, and $\eta=0.243(4)$ for $x=0$ and
$T_c=0.13906(3)$, $1/\nu=1.001(1)$, and $\eta=0.2545(8)$ for $x=0.8$.
The estimates of $\nu$ and $\eta$ are close to those in the two-dimensional Ising universality class, $\nu=1$ and $\eta=1/4$.
The scaling analysis indicates that the phase transitions belong to the two-dimensional Ising universality class.

Carrying out the similar analyses while varying $x$, we obtain the MC phase diagram, as indicated by the red circles in Fig.~\ref{fig:phase}.
The $x$ dependence of $T_c$ is similar to that in the MF result, while the values are reduced to roughly half due to thermal fluctuations.
Interestingly, the scaling collapse works well for all $x$, indicating that the phase transition is always of second order in the MC results. 
This is in clear contrast to the MF result, where both first- and second-order transitions appear with the tricritical point. 
The results suggest that the tricritical behavior and the first-order transition are the artifact of the MF approximation.
In addition to the phase transition, we show the crossover temperature $T^*$ by the orange squares in Fig.~\ref{fig:phase}, which are estimated from the peak $T$ in the specific heat.
Near $x=1$, $T^*$ well agree with $\tilde{T}^*\simeq 0.8336x$ (dashed-dotted line) determined by the peak $T$ of $\tilde{C}_v$ for the two-level system.
With decreasing $x$ from 1, the crossover at $T^*$ merges into the phase transition at $T_c$, and $\tilde{W}_h$ and $\tilde{W}_t$ grows simultaneously in the smaller $x$ region.

\subsection{Comparison to the previous study}

Finally, let us compare the present results with those for the original Hamiltonian given in Eq.~(\ref{eq:1}), which were recently obtained by the authors~\cite{Nasu2015}.
The present results in the limit of $x\to 1$ well explain the asymptotic behaviors of $T_c$ and $T^*$ in the limit of $J' \gg J$ for the original model.
Moreover, the scaling analysis for much larger system sizes compared to the previous study provides compelling evidence of the second-order phase transition belonging to the two-dimensional Ising universality class in the limit of $J'/J\gg 1$.
We note that tricritical behavior was found in the previous study~\cite{Nasu2015}, and it looks qualitatively similar to the MF results in the present study. 
However, this will be just a coincidence, as the tricritical point in the original model is located far from the limit of $J'\gg J$, close to the border between topologically-trivial and nontrivial CSL phases. 
This is far beyond the effective model used in the present analysis. 
Moreover, the tricritical behavior in the effective model is an artifact of the MF approximation as discussed above. 
Further study of more sophisticated effective models including higher-order perturbations will be necessary to clarify the origin of the tricritical behavior. 
This is left for future study.

\section{Conclusion}\label{sec:conclusion}

In summary, we have investigated thermodynamic properties in the isolated dimer limit of the Kitaev-type model on the decorated honeycomb lattice.
Using the MF approximation and the unbiased MC simulation, we found that the effective model exhibits a phase transition from paramagnet to the CSL at a finite $T$.
The MC results indicate that the phase transition is always of second order, whereas the MF approximation gives both first- and second-order transitions with a tricritical point.
The phase transition is associated with the time reversal symmetry breaking, which is identified by the conserved quantities $\tilde{W}_t$.
In addition to the transition, we found a crossover originating from the coherent growth of the other conserved quantities $\tilde{W}_h$.
With increasing the inter-dimer coupling, the crossover is merged into the phase transition.
Moreover, the finite-size scaling analysis shows that the phase transition belongs to the two-dimensional Ising universality class.
We discussed the results in comparison with those for the original Kitaev-type model.


\paragraph{Acknowledgments.} We thank M. Udagawa for fruitful discussion. This work is supported by Grant-in-Aid for Scientific Research No.~15K13533, the Strategic Programs for Innovative Research (SPIRE), MEXT, and the Computational Materials Science Initiative (CMSI), Japan.

%
\label{sect:bib}
\bibliographystyle{unsrt}
\bibliography{easychair}



\end{document}